\documentclass[12pt,preprint]{aastex}

\newcommand{\HII}{H {\small II}  }
\newcommand{\kms}{{\rm ~km~s}^{-1}}
\newcommand{\Msun} {M_\sun}

\slugcomment{To appear in The Astrophysical Journal Letters}

\shorttitle{TIME VARIATION IN G24.78+0.08 A1}
\shortauthors{GALV\'AN-MADRID ET AL.}

\begin{document}


\title{TIME VARIATION IN G24.78+0.08 A1: EVIDENCE FOR AN ACCRETING HYPERCOMPACT \HII REGION?}


\author{Roberto Galv\'an-Madrid\altaffilmark{1,2}, Luis F. Rodr{\'\i}guez\altaffilmark{1}, 
Paul T. P. Ho\altaffilmark{2,3}, and Eric Keto\altaffilmark{2}}


\altaffiltext{1}{Centro de Radioastronom{\'\i}a y Astrof{\'\i}sica, UNAM, Morelia 58090, M\'exico; 
l.rodriguez@astrosmo.unam.mx}
\altaffiltext{2}{Harvard-Smithsonian Center for Astrophysics, 60 Garden Street, 
Cambridge MA 02138, USA; rgalvan, eketo@cfa.harvard.edu}
\altaffiltext{3}{Academia Sinica Institute of Astronomy and Astrophysics, 
Taipei, Taiwan; pho@asiaa.sinica.edu.tw}


\begin{abstract}

Over a timescale of a few years, an observed change in
the optically thick radio continuum flux can indicate whether an unresolved 
\HII region around a newly formed massive star is changing in size. 
In this Letter we report on a study of archival VLA observations 
of the hypercompact \HII region G24.78+0.08 A1 that shows a decrease 
of $\sim 45$ $\%$ in the 6-cm flux over a five year period.
Such a decrease indicates a contraction of $\sim 25$ $\%$ in the ionized 
radius and could be caused by an increase in the ionized gas density if 
the size of the \HII region is determined by a balance between photoionization 
and recombination. 
This finding is not compatible with continuous expansion of the \HII region after the 
end of accretion onto the ionizing star, 
but is consistent with the hypothesis of gravitational trapping and ionized
accretion flows if the mass-accretion rate is not steady. 

\end{abstract}

\keywords{
\HII regions --- ISM: individual (G24.78+0.08) --- stars: formation
}

\section{Introduction} \label{intro}

The formation of massive stars ($M>8$ $\Msun$) by accretion presents a number of theoretical 
difficulties; among them, that once a star attains a sufficient mass 
its surface temperature is high enough to produce a small \HII region 
(see the reviews on ultracompact and hypercompact \HII regions by Churchwell 2002; 
Kurtz 2005; and Hoare et al. 2007). 
The thermal pressure differential between the hot  
($\sim 10^4$ K) ionized gas and the cold ($\sim 100$ K) molecular gas can potentially 
reverse the accretion flow of molecular gas and 
prevent the star from ever reaching a higher mass. However, recent models that include 
the effects of gravity \citep[e.g.,][]{Keto07} have shown 
that if the \HII region is small enough (ionized diameter $\leq 1000$ AU for total stellar 
mass $\geq 350$ $\Msun$), the gravitational attraction of the star(s) 
dominates the thermal pressure and the molecular 
accretion flow can cross the ionization boundary and proceed toward the star as an
ionized accretion flow within the \HII region. In this stage the \HII region is said to be
gravitationally trapped by the star.

Previous studies have demonstrated a number of observational techniques bearing on the
evolution of small \HII regions around newly formed stars.
Accretion flows onto and through \HII regions can be directly observed
by mapping molecular and radio recombination lines (RRLs) at very high angular
resolution
\citep[e.g.,][]{Sol05, KW06}. If the \HII region is too small to be spatially resolved, 
the frequency dependence of the  velocities and widths of 
RRLs can be used to infer steep density gradients and supersonic velocities
within the \HII regions 
\citep{KZK07}, as expected from the presence of ionized accretion flows or bipolar outflows. 
Radio continuum observations 
at very high
angular resolution, made at two epochs some years apart, 
can also be used to directly observe
changes in size of small \HII regions 
(Franco-Hernandez \& Rodriguez 2004). Changes in size indicate
whether
the evolution of the \HII region is consistent with pressure-driven expansion or
gravitational trapping.

In this Letter we demonstrate yet another technique -- how a comparison 
of radio continuum observations 
at different epochs can be used to infer size changes in \HII regions even if the observations
do not spatially resolve them. At optically thick frequencies the radio continuum
flux depends to first approximation only on the size of an \HII region and is independent of its
internal density structure. Therefore, even if the size is not known, a change in the flux over
time still indicates a change in the size of the \HII region.
This technique is particularly valuable because it
can be used on the smallest and youngest \HII regions,
and because lower angular resolution radio continuum observations generally
require less observing time than spectral line and high angular resolution observations.

For the study presented in this Letter we selected the hypercompact (HC) \HII region G24.78+0.08 A1
which lies at the center of a massive molecular accretion flow \citep{Bel04, Bel06} and also has 
multi-epoch 6 cm radio continuum observations in the VLA archive. 
G24.78+0.08 was detected in the centimetric (cm) continuum by \cite{Beck94}, 
and later resolved by \cite{Cod97} into a compact (A) and an extended (B) component. 
A millimeter (mm) interferometric study \citep[]{Bel04} 
revealed the presence of two massive rotating toroids centered in respective dust cores (A1 and 
A2). The compact cm emission comes from the mm component A1 (hereafter G24 A1), and has recently been 
resolved 
by \cite{Bel07}. If G24 A1 is ionized by a single star, its spectral type should be earlier 
than O9 \citep[]{Cod97}. Also, G24 A1 
likely powers a massive CO outflow \citep[]{Fur02}. 

The infalling and rotating molecular gas and  bipolar outflow all
suggest ongoing accretion. However, based on proper motions of H$_2$O masers 
around the \HII region,
\cite{Bel07} and \cite{Mos07} proposed that at the present time, the \HII region is
expanding into the accretion flow.  The suggested timescale for the expansion is short enough that
we should be able to detect a corresponding increase in the optically thick radio continuum
flux within a few years.

\section{Observations} \label{obs}

We searched the VLA\footnote{The 
National Radio Astronomy Observatory is operated by Associated Universities, 
Inc., under cooperative agreement with the National Science Foundation.} 
archive for multi-epoch observations centered in the G24 A1 region 
at optically 
thick frequencies (below 23 GHz for G24 A1). 
Since in the optically thick part of the spectrum the flux density of the source 
scales as the angular size squared, 
flux density variations corresponding to size variations
can be detected even in observations of modest angular resolution. 
We chose three data sets of 6-cm observations 
in the C configuration, two from 1984 and one from 1989 (see Table 1).

The observations were made in both circular polarizations with an effective 
bandwidth of 100 MHz. The amplitude scale was derived from observations of the absolute 
amplitude calibrator 3C286. This scale was transferred to the phase calibrator and
then to the source. We estimate an error not greater than $10$ $\%$ for the flux densities
of the sources.

We edited and calibrated each epoch separately following the standard VLA 
procedures using the reduction software AIPS. Precession to J2000 coordinates was performed running
the task UVFIX in the $(u, v)$ data. After self-calibration, we made CLEANed images 
with uniform weighting and cutting the short spacings (up to 10 K$\lambda$) to minimize the 
presence of extended emission at scales larger than $\sim 20''$.  

Before subtraction, we made the images as similar as possible. We restored the CLEAN
components with an identical Gaussian beam 
HPBW $4\rlap.{''}79 \times 3\rlap.{''}38$, PA$=-15^\circ$.
We applied primary beam corrections and 
aligned the maps. No significant differences were found in the difference image
between the 1984 May 11 and 14 images, as expected for such a small time baseline.
We therefore averaged the two 1984 epochs into a single data set. 
In subtracting the final maps we allowed the 1984 image to have small ($\simeq 0.1''$) 
shifts in position as well as a scaling of $\simeq 10$ $\%$ in amplitude. 
This was done in order to minimize the rms residuals of the difference image in the 
region of interest. A similar procedure was used by \cite{FH04} to detect a variation in the 
lobes of the bipolar UC \HII region NGC 7538 IRS1. 
The individual maps, as well as the final difference image between 1984 and 1989 
are shown in Figure 1.

\section{Discussion} \label{discu}

\subsection{The Expected Variation Trend} \label{trend}

\subsubsection{If G24 A1 Is Expanding} \label{trend-exp}

\citet{Bel07} suggested that the 7 mm and 1.3 cm continuum morphologies
are consistent with limb-brightening from a thin, ionized-shell structure.
Based on H$_2$O maser proper motions
\citep[see Fig. 3 of][ or Fig. 4 of Moscadelli et al. 2007]{Bel07} they 
also suggested an expansion speed of $\sim 40\kms$. 
The increase in flux corresponding to the increase in size due to
expansion ought to be detectable in a few years.

\subsubsection{If G24 A1 Is Accreting} \label{trend-acc}

If G24 A1 is the ionized inner portion of the star forming accretion flow,
the long-term growth of the \HII region due to the increasing
ionizing flux of the star should be imperceptible. However, the \HII region could
change in size over an observable timescale if the gas density in the accretion
flow is time variable. Because the mass of ionized gas within the \HII region
is very small compared to the mass of the accretion flow, even a small
change in the flow density could affect the size of the \HII region.

For example, we can obtain a lower limit for the mass of G24 A1 assuming that 
it is spherical and homogeneous. 
From the equations of \cite{MH67} we have that: 

\begin{eqnarray}
\biggl[\frac{M_{\HII}}{\Msun}\biggr]=3.7\times10^{-5}\biggl[\frac{S_\nu}{\mathrm{mJy}}\biggr]^{0.5} 
& &
\biggl[\frac{T_e}{10^4~\mathrm{K}}\biggr]^{0.175} \times \nonumber \\ 
\biggl[\frac{\nu}{4.9~\mathrm{GHz}}\biggr]^{0.05}
\biggl[\frac{D}{\mathrm{kpc}}\biggr]^{2.5} \biggl[\frac{\theta_s}{\mathrm{arcsec}}\biggr]^{1.5}, 
\nonumber
\end{eqnarray}

\noindent
where the flux density $S_\nu$ has been measured at a frequency $\nu$ in which the \HII region is 
optically thin, $T_e$ is the electron temperature, $D$ is the distance to the region, and $\theta_s$ 
 is its FWHP. Taking $S_\nu=101$ mJy, $D=7.7$ kpc, and $\theta_s=0\rlap.{''}17$ at 7 mm   
\citep[]{Bel07}, and assuming $T_e=10^4$ K, the ionized mass in G24 A1 would be 
$\sim 5 \times 10^{-3}\Msun$. A higher limit to the mass can be obtained 
considering that HC \HII regions should have density 
gradients ($n \propto r^\alpha$, with $\alpha=-1.5$ to $-2.5$) rather than being homogeneous 
\citep[]{Keto07}. In this case we obtain ionized masses between $1 \times 10^{-2}$ $\Msun$ and 
$3 \times 10^{-2}$ $\Msun$, a factor of $2-6$ higher than under the assumption of homogeneity, but still 
very small when compared to that of the surrounding molecular material or the ionizing star itself.

\subsection{The Observed Variation} \label{obsvar}

Figure 1 shows the individual maps of the 1984 May $11+14$ (1984.36) and 1989 June 23 
(1989.48) epochs, as well as the difference image. Sources A1 and B \citep[the former 
labeled A by][]{Cod97} 
are unresolved ($\theta_s \leq 2{''}$), and the total emission is dominated by component B. 
However, the 
flux decrease is centered at the position of A1. We performed Gaussian fits to both 
components using the task JMFIT in AIPS, 
and the results are summarized in Table 2. 
We have checked the reliability of the fits to the image by
making direct fits to the {\it (u,v)} data. The values obtained
from both techniques are entirely consistent but suggest that
the errors given by the tasks used in the image (JMFIT) and
{\it (u,v)} (UVFIT) fittings are underestimated by a factor 
of 2. The errors given in Table 2 have been corrected by this factor. 

The decrease in the flux density of G24 A1 
between 1989.48 and 1984.36 is  $5.1 \pm 1.1$ mJy, or $45 \pm 10$ $\%$. Component B shows no 
evidence of time variability, as expected for a more evolved \HII region not associated 
with signs of current star-forming activity such as H$_2$O or OH masers \citep[][]{Cod97}. 
We set an upper limit of 2 $\%$ to the circularly-polarized emission of the sources, which 
indicates that we are not dealing with variable gyrosynchrotron emission from an 
active stellar magnetosphere.

\subsection{Is G24 A1 Accreting?}

The $\sim 45$ $\%$ flux decrease (i.e., a contraction of $\sim 25$ $\%$ in the ionized radius) 
we have detected at 6 cm toward G24 A1 is not consistent with the hypothesis that this \HII 
region is expanding rapidly into the molecular accretion flow. 
Assuming a radius $R \sim 500$ AU and an expansion velocity $v \sim 40 \kms$ \citep{Bel07}, a  
flux {\it increase} of $\sim 20$ $\%$ should have been observed between the compared epochs. 

We attribute the contraction of the \HII region to an increase in its density produced by the 
enhancement of accretion, either caused by an isotropic increment in the mass-accretion rate or 
by the sudden accretion of a localized clump in the neutral inflow 
\citep[for example, clumps have been observed in the accretion flow onto the 
UC \HII region G10.6--0.4;][]{SoHo05,KW06}. 
Our data do not allow us to distinguish among these two possibilities. 

The additional mass required to increase the density can be estimated.
The ionized radius $r_{\rm s}$ scales roughly with the density $n$ as 
$r_{\rm s} \propto n^{-2/3}$ (if the radius of the \HII region is set by a balance 
between photoionization and recombination) and at optically thick frequencies the 
flux density scales as $S_\nu \propto r_{\rm s}^2$. 
The ionized mass within G24 A1 is 
$\sim 1 \times 10^{-2}$ $\Msun$ (\S \ref{trend-acc}). Therefore,  
the sudden accretion of  
$\sim 5 \times 10^{-3}$ $\Msun$ into the \HII region would suffice to explain 
the observed $45$ $\%$ flux decrease. 
The molecular accretion flow
around this \HII region has a mass of $\sim 130$ $\Msun$ \citep[][]{Bel04}.
Thus, the required variation, 50 $\%$ of the mass of the \HII region, is 
only 0.001 \% of the total mass of the accretion flow.

Finally, we shall mention that even when the observations here reported are not 
consistent with a simple, continuous expansion of G24 A1, they do not rule out 
other more complex, non-steady evolutionary scenarios. 

\section{Conclusions}

Our analysis of archival VLA observations of the \HII region G24.78+0.08 A1 
indicates a contraction of its radius between 1984.36 and 1989.48. This 
finding is consistent with the hypothesis that this HC \HII is the inner, ionized part of the 
larger scale accretion flow seen in the molecular line observations of \citet{Bel04,Bel06}.  
Future high-resolution RRL observations could unambiguously resolve the ionized-gas 
kinematics and confirm this hypothesis.

\clearpage

\begin{center} 
\begin{deluxetable}{ccccccc}  \label{tab1}
\tabletypesize{\scriptsize}
\tablecaption{Observational Parameters}
\tablewidth{0pt}
\tablehead{
\colhead{Epoch} & \multicolumn{2}{c}{Phase Center\tablenotemark{a}} & \colhead{Amplitude} & 
\colhead{Phase} & 
\colhead{Bootstrapped Flux} & \colhead{Beam}  \\\cline{2-3}
 & $\alpha$(J2000) & $\delta$(J2000) & Calibrator & Calibrator & Density (Jy) &(arcsec $\times$ 
arcsec; deg)  \\
}
\startdata
1984 May 11 & 18 36 12.145 & $-07$ 11 28.17 & 3C286 & $1743-038$ & $2.414 \pm 0.005$ & 
$5.38 \times 3.44$; $-29$  \\
\\
1984 May 14 & 18 36 12.145 & $-07$ 11 28.17 & 3C286 & $1743-038$ & $2.464 \pm 0.007$ & 
$4.78 \times 3.34$; $+03$  \\
\\
1989 Jun 23 & 18 36 10.682 & $-07$ 11 19.87 & 3C286 & $1834-126$ & $0.153 \pm 0.001$ & 
$4.50 \times 3.31$; $-10$   

\enddata
\tablenotetext{a}{Units of right ascension are hours, minutes, and seconds. Units of declination are 
degrees, arcminutes, and arcseconds.}
\end{deluxetable}
\end{center}

\clearpage

\begin{center} 
\begin{deluxetable}{ccccc}  \label{tab2}
\tabletypesize{\scriptsize}
\tablecaption{Gaussian Fit Results}
\tablewidth{0pt}
\tablehead{
\colhead{Epoch} & \colhead{Component} & \multicolumn{2}{c}{Position\tablenotemark{a,b}} & 
\colhead{Flux Density} 
\\\cline{3-4}
 & & $\alpha$(J2000) & $\delta$(J2000) & (mJy) \\
}
\startdata
1984.36 & A1 & 18 36 12.545 & $-07$ 12 10.87 & $11.4 \pm 0.8$ \\
1984.36 & B  & 18 36 12.668 & $-07$ 12 15.37 & $31.4 \pm 1.0$ \\
1989.48 & A1 & 18 36 12.545 & $-07$ 12 10.87 & $06.3 \pm 0.8$ \\
1989.48 & B  & 18 36 12.668 & $-07$ 12 15.37 & $32.8 \pm 0.8$ \\
\enddata
\tablenotetext{a}{Units of right ascension are hours, minutes, and seconds. Units of declination are 
degrees, arcminutes, and arcseconds.}
\tablenotetext{b}{For the 1989.48 epoch, the centers of the Gaussians were fixed to those obtained for 
the 1984.36 epoch.}
\end{deluxetable}
\end{center}

\clearpage

\begin{figure}
\epsscale{.30}
\plotone{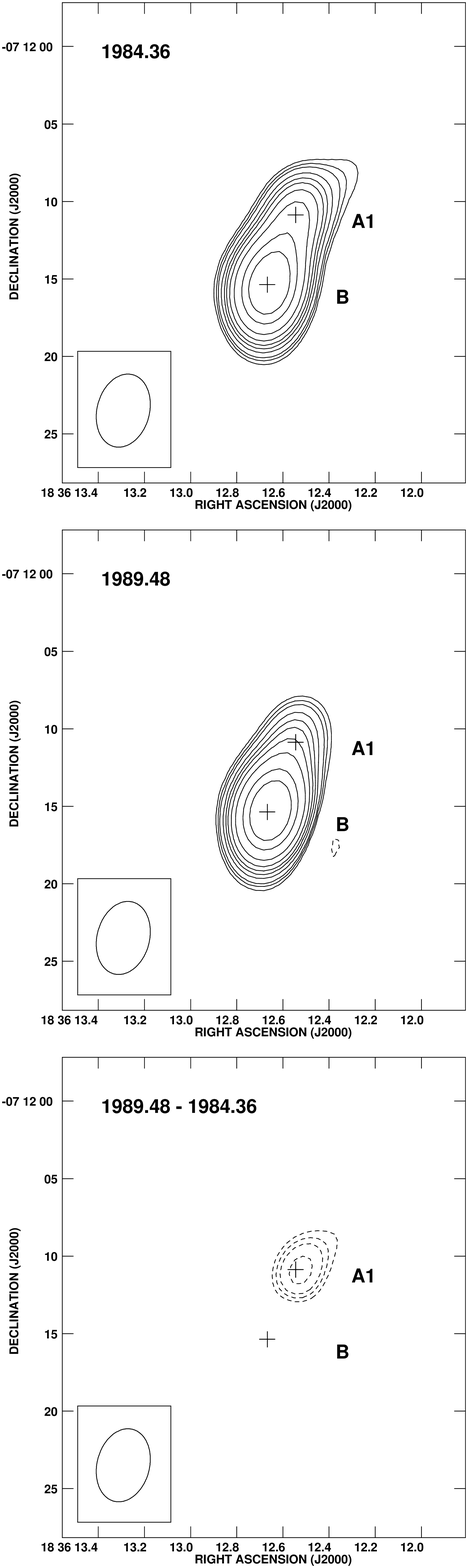}
\caption{
VLA images of G24.78+0.08 for 1984.36 ({\it top}), 1989.48 ({\it middle}), and the difference 
of 1989.48 - 1984.36 ({\it bottom}). The contours are -10, -8, -6, -5, -4, 4,
5, 6, 8, 10, 12, 15, 20, 30, 40, and 60 times 0.57 mJy beam$^{-1}$. 
The half power contour of the synthesized beam ($4\rlap.{''}79 \times
3\rlap.{''}38$ with a position angle of $-15^\circ$) is shown in the bottom left corner 
of the images.
The crosses indicate the positions of the components A1 and B from our Gaussian fits to the 
1984.36 image. The negative residuals observed in the difference image indicate a decrease 
of $\sim 45$ $\%$
in the flux density of component A1. 
}
\label{fig1}
\end{figure}




\end{document}